\documentclass[floatfix,aps,eqsecnum,prd,showpacs,amsmath,nofootinbib,
preprintnumbers,twocolumn]{revtex4}

\usepackage{graphicx}

\newcommand{\beq}{\begin{equation}}
\newcommand{\eeq}{\end{equation}}
\newcommand{\bea}{\begin{eqnarray}}
\newcommand{\eea}{\end{eqnarray}}

\def\laq{~\raise 0.4ex\hbox{$<$}\kern -0.8em\lower 0.62
ex\hbox{$\sim$}~}
\def\gaq{~\raise 0.4ex\hbox{$>$}\kern -0.7em\lower 0.62
ex\hbox{$\sim$}~}

\def \pa {\partial}
\def \ra {\rightarrow}

\def \Da {\Delta}
\def \b {\beta}
\def \a {\alpha}

\def \Ga {\Gamma}
\def \ga {\gamma}

\def \da {\delta}
\def \ep {\epsilon}
\def \r {\rho}

\def \Om {\Omega}

\def \u {\hat u}
\def \v {\hat v}
\def \n{\hat n}
\def \x{\hat x}
\def \y{\hat y}
\def \bm {\overline m}

\begin{document}

\preprint{BA-TH/05-511}
\preprint{gr-qc/0504079}

\title{Response of the common mode of interferometric detectors \\
to a stochastic background of massive scalar radiation }
\author{N. Bonasia and M. Gasperini}\email{gasperini@ba.infn.it}
\affiliation{Dipartimento di Fisica, Universit\`a di Bari, 
Via G. Amendola 173, 70126 Bari, Italy \\
and Istituto Nazionale di Fisica Nucleare, Sezione di Bari,
Bari, Italy}


\begin{abstract}
We compute the angular pattern and the overlap reduction functions for the geodesic and non-geodesic response of the common mode of two interferometers interacting with a stochastic, massive scalar background. We also discuss the possible overlap between common and differential modes. We find that the 
cross-correlated response of two common modes to a non-relativistic background may be higher than the response of two differential modes to the same background. 
\end{abstract}


\maketitle

\section {Introduction}
\label{I}
It is well known that interferometric detectors can respond to the gravitational radiation in two ways, either through the ``differential" or the ``common" mode \cite{1,2}, described, respectively, by the two response tensors $D_-^{ij}$ and $D_+^{ij}$,
\beq
D_\pm^{ij}= \u^i\u^j\pm \v^i\v^j, 
\label{11}
\eeq
where $\u$, $\v$ are unit vectors aligned along the arms of the interferometer. All theoretical  and phenomenological analyses presently available in the literature, however, are almost exclusively devoted to the response of the differential mode. In the case of tensor (spin-two) gravitational radiation there is indeed no need of considering the common mode, which is expected to be much more  ``noisy" (from the point of view of the experimental efficiency), and which is ``blind" to the so-colled ``cross" polarization state of tensor gravitational radiation. 

According to present schemes of high-energy unification of all interactions (such as supergravity and superstring models), however, the graviton is expected to have scalar partners. As a consequence, the cosmic background of gravitational waves of primordial (inflationary) origin could be associated to a relic background of scalar, possibly massive, particles \cite{3}, interacting (with nearly gravitational strength) with existing gravitational antennas. The antenna response to massive scalar radiation differs in a significant way from the response to  tensor radiation, mainly because of the different polarization states of the scalar particles with respect to the gravitons. 

Previous studies on the interaction of massive scalar waves with interferometric detectors were all concentrated on the  differential mode \cite{4,5,6}. The aim of this paper is to discuss the response of the common mode to a stochastic background of massive, scalar radiation. It will be shown, in particular, that the common mode can interact with scalars as efficiently as the differential mode and that, in the case of non-relativistic radiation, the correlated signal of two detectors may be enhanced (in some cases) for the common mode with respect to the differential mode, at fixed values of  all the other observational parameters. Such an enhancement could (in principle) compensate the higher level of noise expected in the context of a data analysis based on the common mode response. 

The paper is organized as follows. In Sect. II we compute the so-called antenna pattern functions for the response of the common mode to scalar and tensor radiation, comparing them with the (known) results relative to the differential mode. In Sect. III we compute the overlap reduction function for the cross-correlation of the common modes of two interferometers, while in Sect. IV we compute the overlap function for the cross-correlation of common and differential modes. In Sect. V we discuss a possible enhancement of the signal-to noise ratio for non-relativistic spectra. The main results of this paper are finally summarized in Sect. VI. 

\section{Pattern functions for scalar-tensor signals}
\label{II}

It is well known that the detection of a stochastic background of gravitational radiation requires the cross-correlation of the outputs of at least two detectors \cite{1,2}. The analysis of such a correlation requires the knwledge of the signal, or ``physical strain" $h(t)$, induced by the incident radiation on the single detector. Such a signal is obtained through the projection, on the  arms of the detector, of the ``tidal" forces which are due to the incident radiation, and which are described by the equation of geodesic deviation (see for instance \cite{7}). 

For a test particle moving on a generic scalar-tensor background, the equation of geodesic deviation deviation has to be generalized to take into account the possible direct coupling of the particles to the gradients of the scalar background $\phi$ (in addition to the gradients of the metric), and is given by \cite{8}
\beq
{D^2 \eta^\mu \over D\tau^2} +\eta^\b R_{\b\a\nu}\,^\mu u^\a u^\b +
q\eta^\b \nabla_\b\nabla^\mu \phi =0.
\label{21}
\eeq
Here $\eta^\mu$ is the infinitesimal spacelike vector connecting two nearby (non-geodesic) trajectories, and $q$ is the (dimensionless) scalar charge per unit of gravitational mass of the given test particles. Perturbing the metric background in the weak field approximation, $g_{\mu\nu}= \eta_{\mu\nu} + \da g_{\mu\nu}$, and taking into account the possible scalar components of the metric fluctuations induced by the scalar field fluctuations $\da \phi$, we can set, in the longitudinal gauge \cite{9}:
\beq
\da g_{\mu\nu}= {\rm diag} \left( 2 \Phi, h_{ij} +2 \Psi \da_{ij} \right),
~~~~~ \da \phi= X.
\label{22}
\eeq
Here $h_{ij}$ is the transverse and traceless (pure tensor) part of metric fluctuations, $\Phi$ and $\Psi$ are the gauge-invariant Bardeen potentials, and $X$ is the gauge -invariant scalar field perturbation. Conventions: $g_{00}>0$, $R_{\nu\a}=R_{\mu\nu\a}\,^\mu$, and 
$R_{\mu\nu\a}\,^\b=\pa_\mu\Ga_{\nu\a}\,^\b+ \Ga_{\mu \rho}\,^\b
\Ga_{\nu \a}\,^\rho -...$). In the non-relativistic and weak field limit, the relative acceleration of two test particles with proper separation $L^k$, according to Eq. (\ref{21}), is then given by
\bea
&&
\ddot \xi^i= - L^k M_k^i, ~~~ M_{ij}= \da R_{iooj} + q \pa_i \pa_j X =
\nonumber\\
&&
=-{1\over 2} \ddot h_{ij} +  \pa_i \pa_j \Phi - \da_{ij} \ddot \Psi + 
q  \pa_i \pa_j X,
\label{23}
\eea
where $M_{ij}$ is the total stress tensor which includes the contribution of scalar and tensor radiation (see also \cite{10} for the case $q=0$). 

For a stochastic background it is now convenient to adopt a Fourier expansion of the fluctuations. Tensor fluctuations are massless, and can be expanded into frequency modes $h_{ij}(\nu, \n)$, where $\n$ is a unit vector specifying the propagation direction; scalar fluctuations are possibly massive, and should be expanded into momentum modes $X(p, \n)$, with frequency $\nu= E(p)= (p^2+ \bm^2)^{1/2}$, where $\bm= m/2 \pi$ (note that we are using ``unconventional" units $h=1$, i.e. $\hbar =1/2 \pi$, for an easier comparison with experimental variables). For a unified treatment we set $\nu=E=p$ for massless radiation, and we expand tensor and scalar fields as
\bea
&&
h_{ij}= {1\over 2} \int _{-\infty}^{+\infty} dp \int_{\Om_2} d^2 \n~
\ep^A_{ij} h_A(p, \n) e^{2\pi i p (\n \cdot x -t)} + h.c. \nonumber 
\\
&&
X= {1\over 2} \int  _{-\infty}^{+\infty} dp \int_{\Om_2} d^2 \n~
X(p, \n) e^{2\pi i (p \n \cdot x -Et)}\nonumber \\
&&
~~~~~~ + h.c.
\label{24}
\eea
(a similar expansion holds for $\Psi$ and $\Phi$). The tensor $\ep^A_{ij}$, $A=1,2$, describes the transverse and traceless polarization states of spin-two gravitational radiation, and the angular integration $d^2 \n$ extends over the full solid angle $\Om_2$. The total tidal tensor thus becomes
\bea
&&
M_{ij}=  {1\over 2} \int _{-\infty}^{+\infty} dp \int_{\Om_2} d^2 \n~
(2 \pi E)^2 \Bigg[ {1\over 2} \ep^A_{ij}h_A + \da_{ij} \Psi 
\nonumber \\
&&
\qquad 
-\left(1-{\bm^2\over E^2} \right) \n_i\n_j \Phi - q{p^2 \over E^2}  \n_i\n_j 
X \Bigg] e^{2\pi i (p \n \cdot x -Et)}\nonumber \\
&&
~~~~~~ + h.c.
\label{25}
\eea

In the absence of sources for scalar fluctuations other than the scalar field $X$ we can set $\Phi=\Psi$ \cite{9}, and the scalar contribution to $M_{ij}$ arising from the ``electric" components $R_{i00j}$ of the Riemann tensor can be decomposed as 
\beq
(T_{ij} + {\bm^2 \over E^2} L_{ij}) \Psi,
\label{26}
\eeq
where
\beq
T_{ij}= \da_{ij} - \n_i\n_j, ~~~~~~~
L_{ij}=  \n_i\n_j
\label{27}
\eeq
are, respectively, the transverse and longitudinal components of the scalar stress tensor. We shall refer to such a Riemannian contribution as to the ``geodesic" part of the scalar tidal forces, while the direct contribution arising from the scalar field gradients , $-q (p/E)^2 L_{ij} X$, will be referred to as the ``non-geodesic" part. By setting $ M_{ij}= - \ddot F_{ij}$, the projection over the detector response tensor $D^{ij}_\pm$ leads finally to the physical strain $h=F_{ij}D^{ij}_\pm$, at the detector position $\vec x_0=$ const,
\bea
&&
h(t, \vec x_0)=
{1\over 2} \int _{-\infty}^{+\infty} dp \int_{\Om_2} d^2 \n~\Bigg[
F_\pm^A(\n) h_A(p,\n)+
\nonumber \\
&&
\qquad
+ F_\pm^{\rm geo}(\n)\Psi (p,\n)
 +  F_\pm^{\rm ng}(\n)X(p,\n) \Bigg]
e^{2\pi i (p \n \cdot \vec x_0 -Et)} 
\nonumber \\
&&
~~~~~~~~~~~   + h.c. ,
\label{28}
\eea
where 
\bea
&&
F_\pm^A={1\over 2} D_\pm^{ij} \ep^A_{ij}, ~~~~~~~
F_\pm^{\rm geo}= D_\pm^{ij} (T_{ij} + {\bm^2 \over E^2} L_{ij}), 
\nonumber \\
&&
F_\pm^{\rm ng}= -q {p^2\over E^2} D_\pm^{ij} L_{ij},
\label{29}
\eea
are the tensor and scalar part of the antenna pattern functions \cite{6}. The plus and minus sign refers to the common and differential mode, respectively, defined in Eq. (\ref{11}). It should be noticed that the longitudinal scalar strain survives even in the massless limit, because of the presence of the direct coupling to the gradients of the scalar background, $\pa_i\pa_j X$. 

For a graphic illustration of the various pattern functions we may consider a convenient frame in which the two arms of the interferometer, represented by the unit vectors $\u$ and $\v$, are coaligned with the $x_1$ and $x_2$ axes, respectively, and the direction $\n$ of wave propagation is specified by the polar and azimuthal angles $\phi$ and $\theta$:
\bea
&&
\u=(1,0,0), ~~~~~ \v=(0,1,0), \nonumber \\
&&
\n= (\sin \theta \cos \phi, \sin \theta \sin \phi, \cos \theta).
\label{210}
\eea
We also introduce, in such a frame, two unit vectors $\x$ and $\y$, 
orthogonal to $\n$ and to each other,
\bea
&&
\x=(\sin \phi, -\cos \phi,0), \nonumber \\
&&
 \y= (\cos \theta \cos \phi, \cos \theta \sin \phi, -\sin \theta). 
 \label{211}
\eea
In terms of these vectors, the two independent tensor polarization states can then parametrized as
\bea
&&
\ep^{(+)}_{ij}= \x_i\x_j-\y_i\y_j,  \nonumber \\
&&
\ep^{(\times)}_{ij}= \x_i\y_j+\y_i\x_j.
\label{212}
\eea

\begin{figure}[t]
\begin{center}
\includegraphics[width=37mm]{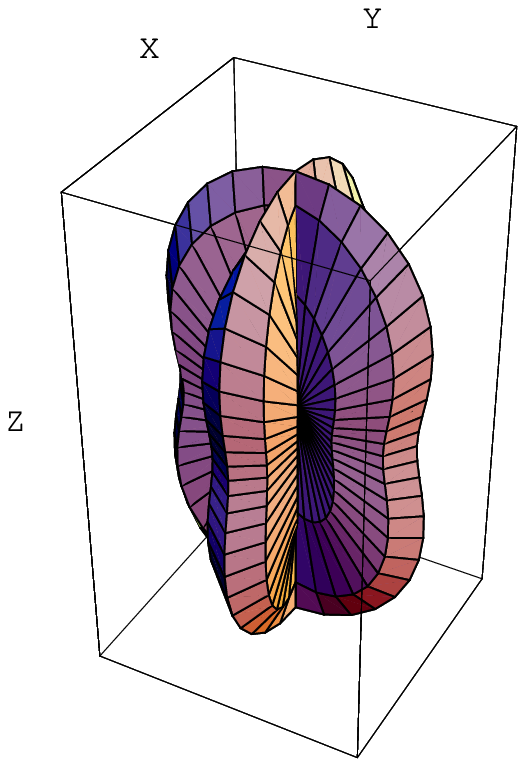}~~~
\includegraphics[width=30mm]{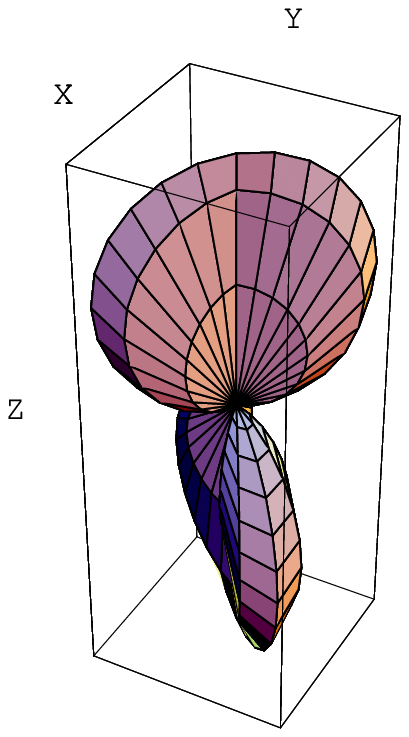}\\
\includegraphics[width=65mm]{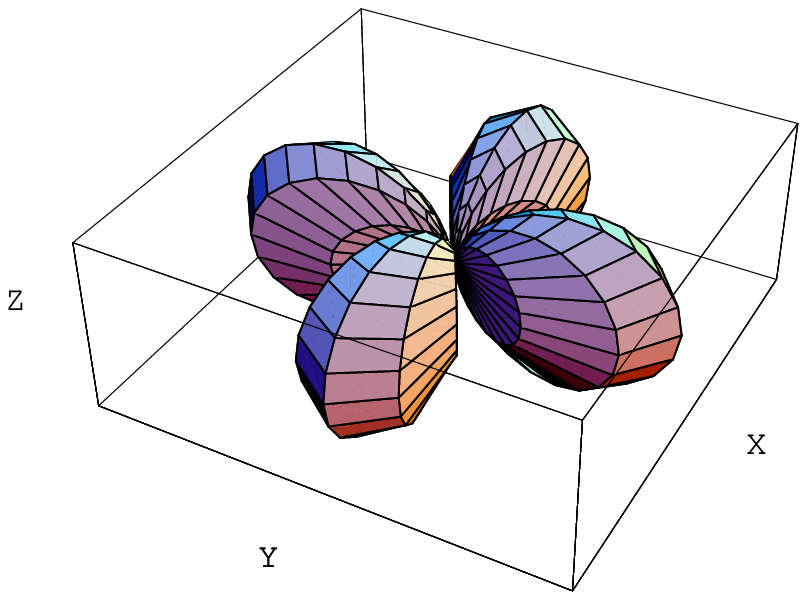}
\end{center}
\caption{\sl Angular pattern functions for the differential mode of an interferometer. The top panel illustrates the response to tensor radiation, $F_-^{(+)}$ (left) and $F_-^{(\times)}$ (right), of Eq.(\ref{213}). The bottom panel illustrates  the response to scalar radiation of Eq. (\ref{214}), describing both the geodesic and non-geodesic pattern (which are proportional).} 
\label{f1}
\end{figure}
For the response of the differential mode we then easily obtain the usual tensor pattern functions \cite{1},
\bea
&&
F_-^{(+)}={1\over 2} D_-^{ij} \ep^{(+)}_{ij}= -{1\over 2} (1+ \cos^2 \theta) \cos 2 \phi,
\nonumber\\
&&
F_-^{(\times)}={1\over 2} D_-^{ij} \ep^{(\times)}_{ij}= \cos \theta \sin 2 \phi,
\label{213}
\eea
and the (already known) scalar pattern functions \cite{4,6},
\beq
F_-^{\rm geo}= -\left( p\over E \right)^2\sin^2 \theta \cos 2 \phi, ~~~~
F_-^{\rm ng}=q F_-^{\rm geo}.
\label{214}
\eeq
The angular distribution of the various pattern functions is  illustrated in Fig. 1. 
In the same way we obtain, for the common mode $D_+^{ij}$, the tensor patter functions, 
\bea
&&
F_+^{(+)}={1\over 2} D_+^{ij} \ep^{(+)}_{ij}= {1\over 2}\sin^2 \theta, 
\nonumber\\
&&
F_+^{(\times)}={1\over 2} D_+^{ij} \ep^{(\times)}_{ij}=0,
\label{215}
\eea
and the new scalar pattern functions,
\bea
&&
F_+^{\rm geo}= 1+ \cos^2 \theta +\left(\bm\over E\right)^2 \sin^2 \theta=2-  \left(p\over E\right)^2\sin^2 \theta,
\nonumber\\
&&
F_+^{\rm ng}=- q \left(p\over E\right)^2\sin^2 \theta,
\label{216}
\eea
which are illustrated in Fig. 2. 

\begin{figure}[t]
\begin{center}
\includegraphics[width=45mm]{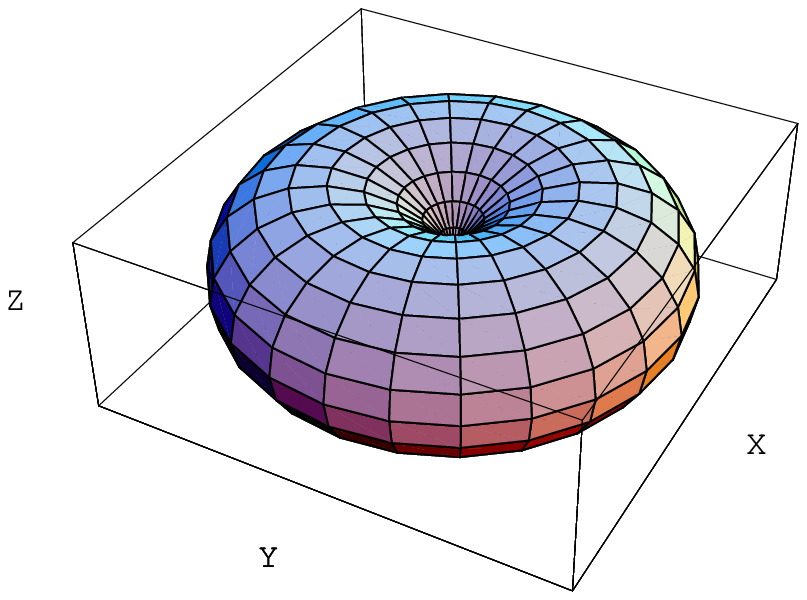}~~
  \includegraphics[width=35mm]{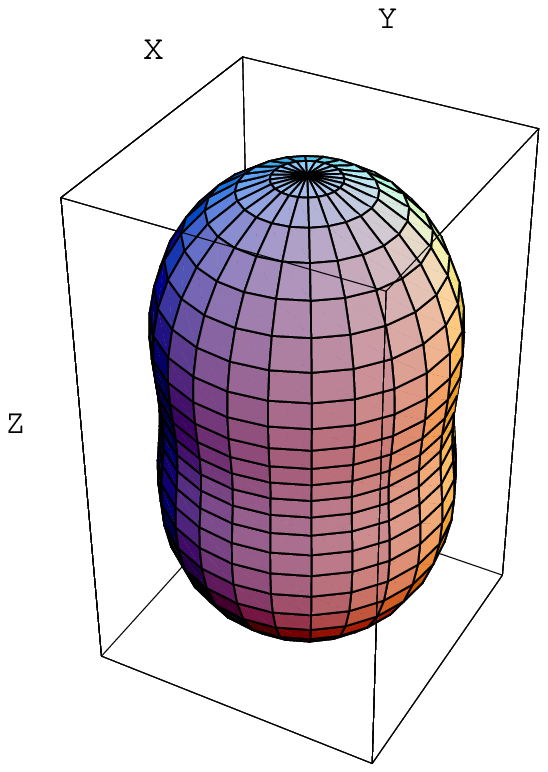}
  \end{center}
\caption{\sl Angular pattern functions for the common mode of an interferometer. The left panel illustrates both the tensor response $F_+^{(+)}$ of Eq.(\ref{215}), and the scalar response  $F_+^{\rm ng}$ of Eq. (\ref{216}). The right panel illustrates the geodesic response $F_+^{\rm geo}$ for relativistic scalar radiation with $\bm/E=1/3$.} 
\label{f2}
\end{figure}

As already stressed in \cite{6} , the response of the differential mode to 
non-relativistic scalar radiation tends to be highly suppressed because of the factor $p/E \ll1$. Such a suppression is absent, however, for the geodesic part of the response of the common mode which, in the extremum limit $p \ra 0$, $ m \ra E$, tends to become isotropic and even maximal, $F_+^{\rm geo} \ra 2$. The important consequences of this property will be discussed in the following sections. 

\section{Overlap between common modes}
\label{III}

The rest of this paper will be devoted to discuss the case of massive scalar radiation. We shall consider the correlated outputs of two independent detectors, $s_i=h_i+n_i$, $i=1,2$, assuming that the noises are statistically uncorrelated, $\langle n_in_j\rangle =0=\langle n_i h_j\rangle$, $i \not=j$, and much larger than the physical strains, $|n_i| \gg |h_i|$. The correlated signal, over a given observation time $T$, is then given by
\bea
&&
\langle S \rangle = \int_{-T/2}^{T/2} dt~ dt'~ \langle s_1(t) s_2(t') \rangle Q(t-t')= \nonumber \\
&&
= \int_{-T/2}^{T/2} dt~ dt'~ \langle h_1(t) h_2(t') \rangle Q(t-t'),
\label{31}
\eea
where $Q$ is an appropriate filter function \cite{1}, chosen in such a way as to optimize the signal-to-noise ratio. The average brackets are referred to the stochastic properties of the scalar backgrounds $X$ and $\Psi$. A stationary, isotropic, unpolarized (massive) background is characterized, in particular, by the following stochastic conditions:
\bea
&&
\langle X(p, \n)\rangle =0, ~~~~~~~~
\langle X(p, \n) X(p', \n')\rangle=0, \nonumber \\
&&
\langle X(p, \n) X^\ast(p', \n')\rangle= \da(p-p') {\da^2(\n,\n')\over 4 \pi} {1\over 2} S_X(|p|)\nonumber \\
&&
\label{32}
\eea
(similar relations hold for the geodesic scalar background $\Psi$). Here $S_X(|p|)$ is the scalar ``strain density", normalized as in the case of tensor fluctuations. 

The strain density can be conveniently expressed in terms of the spectral energy density of the stochastic background, $d \rho/ d \ln p$, starting with the (canonically normalized) energy density of the scalar fluctuations, 
\beq
\rho_X= \langle \tau_0^0 \rangle = {M_P^2 \over 4} \langle 
\dot X^2 +(\pa_i X)^2 + m^2 X^2 \rangle
\label{33}
\eeq
($M_P$ is the Planck mass). Using the expansion (\ref{24}), and the stochastic conditions (\ref{32}), we obtain
\beq
\rho_X=4 \pi^2 {M_P^2 \over 4}  \int_0^\infty dp~S_X(|p|) E^2(p),
\label{34}
\eeq
from which, in units of critical energy density $\r_c = 3 H_0^2 M_P^2$,
\beq
\Om_X= {p\over \r_c} { d \r_X \over dp}= {\pi^2 |p| E^2 \over 3 H_0^2} S_X(|p|),
\label{35}
\eeq
and
\beq
\langle X(p, \n) X^\ast(p', \n')\rangle= \da(p-p') {\da^2(\n,\n')} {3 H_0^2\over 8 \pi^3 |p| E^2} \Om_X(p).
\label{36}
\eeq
In the limit $m \ra 0, p \ra E=\nu$, we may thus recover standard result for a stochastic gravitational wave background \cite{1}, modulo the additional factor $1/4$ present in the tensor case, and due to the fact that the graviton spectrum $\Om_g(p)$ contains, for a given $S_g(|p|)$, the contribution of two polaritazion states, each one of them associated to two possible helicity configurations, $\pm2$. 

The above result can now be applied to the explicit computation of the correlated signal (\ref{31}). For the non-geodesic part of the signal, due to the $X$ background (see Eq. (\ref{28})), we obtain in particular
\bea
&&
\langle S \rangle= {3 H_0^2 \over 32 \pi^3} \int_{-T/2}^{T/2} dt~ dt'~\int_{-\infty}^{+\infty} dE'~ Q(E') \nonumber\\
&&
\times \int_{-\infty}^{+\infty} 
{dp \over |p| E^2} \Om_X(p) 
 \int_{\Om_2} d^2 \n~F_{1 \pm}^{\rm ng}~ F_{2 \pm}^{\rm ng}
e^{2 \pi i p \n \cdot \Da \vec x}\nonumber\\
&&
\times \left[ e^{-2 \pi i(E+E') \Da t} + 
 e^{2 \pi i(E-E') \Da t}\right],
\label{37}
\eea
where  $\Da \vec x = \vec x_1 -\vec x_2$ is the spatial separation vectors of the two detectors, $ \Da t= t-t'$, and $Q(E)$ is the Fourier transform of the filter function,
\beq
Q(t-t')= \int_{-\infty}^{+\infty} dE~Q(E)  e^{-2 \pi iE(t-t')}.
\label{38}
\eeq
Finally, $F_{i \pm}^{\rm ng}(\n)$ are the scalar pattern functions of the two detectors. The time integration can be easily performed by assuming, as in the tensor case, that the observation time $T$ is much larger than the typical time intervals over which $Q \not=0$, and extending to $\pm \infty$ the limits of the $dt'$ integration. This leads to the Dirac deltas $\da (E+E')$ and $\da (E-E')$. Integration over $dE'$ and $dt$ becomes then trivial, and assuming $Q(E)=Q(-E)$ we obtain the final result
\beq
\langle S \rangle= NT{3 H_0^2 \over 16 \pi^3}
\int_{-\infty}^{+\infty} 
{dp \over |p| E^2} \ga_\pm^{\rm ng}(p) Q(E(p)) \Om_X(p),
\label{39}
\eeq
where 
\beq
 \ga_\pm^{\rm ng}(p)= {1\over N} 
 \int_{\Om_2} d^2 \n~F_{1 \pm}^{\rm ng}(\n)~ F_{2 \pm}^{\rm ng}(\n) 
e^{2 \pi i p \n \cdot \Da \vec x}
\label{310}
\eeq
is the so-called overlap reduction function \cite{11} ($N$ is an arbitrary normalization factor). In the case of the geodesic scalar background $\Psi$ one exactly obtains the same expression, with the obvious replacements $\Om_X \ra \Om_\Psi$, $F_i^{\rm ng} \ra F_i^{\rm geo}$, and $ \ga_\pm^{\rm ng} \ra
 \ga_\pm^{\rm geo}$. 

The scalar overlap functions $\ga_-$ have already been  computed in \cite{4} and discussed in \cite{6}. In this section we present the results for the scalar overlap functions 
$ \ga_+^{\rm ng}$, $ \ga_+^{\rm geo}$ associated to the common modes of two interferometers. We shall consider, for simplicity, the case of two co-planar detectors (which is a realistic assumption if the spatial separation $|\Da \vec x|=d$ of the interferometers is much smaller than the Earth curvature radius). The results are independent of the relative orientation of the axes of the detectors, thanks to the rotational symmetry of $F_+$ with respect to the polar angle $\phi$. Using the explicit definitions of $F_+^{\rm ng}$ and $F_+^{\rm geo}$, and using the integral representation of the spherical Bessel functions, the angular integration of Eq. (\ref{310}) leads to
\bea
&&
 \ga_+^{\rm ng}(p)=q_1q_2{4 \pi \over N} \left(p\over E\right)^4 \left[j_0(\a)-{2\over \a} j_1(\a)+{3\over \a^2} j_2(\a)\right],
\nonumber \\
&&
 \ga_+^{\rm geo}(p)={4 \pi \over N} \Bigg[\left(4-4{p^2\over E^2}+{p^4\over E^4}\right)j_0(\a)
\nonumber\\
&&
\qquad
+{1\over \a}\left(4{p^2\over E^2}-2{p^4\over E^4}\right)j_1(\a)
+{3\over \a^2}\left(p\over E\right)^4 j_2(\a)\Bigg],
\nonumber \\
&&
\label{311}
\eea
where $\a=2 \pi p d$, and $j_0,j_1,j_2$ are spherical Bessel functions,
\bea
&&
j_0(\a)= {\sin \a \over \a}, ~~~ j_1(\a) ={ j_0(\a)\over \a}
- \cos \a, 
\nonumber \\
&&
j_{l+1(\a)} = {2 l+1\over \a} j_l(\a)- j_{l-1}(\a), ~~~ l \geq 1.
\label{312}
\eea
It is important to notice, for the further discussion, that the non-geodesic overlap $ \ga_+^{\rm ng}$ goes to zero as $(p/E)^4$ for $p \ra 0$, exactly like the overlap of two differential modes \cite{6}. The geodesic overlap, instead, goes to a constant, 
$\ga_+^{\rm geo}(0)=16 \pi/N$, like in the case of resonant, spherical mass detectors \cite{12}. 

\section{Overlap between common and differential modes}
\label{IV}

It may be useful, for phenomenological applications, to consider also the overlap between the common mode of one interferometer and the differential mode of another (or even of the same) interferometer. The result of such an overlap is strongly dependent on the relative geometric arrangement of the two detectors.

Consider for instance a scalar background geodesically interacting with the detectors, and generating the ``mixed" overlap function 
\beq
 \ga_{+-}^{\rm gep}(p)= {1\over N} 
 \int_{\Om_2} d^2 \n~F_{1 +}^{\rm geo}(\n)~ 
F_{2 -}^{\rm geo}(\n) 
e^{2 \pi i p \n \cdot \Da \vec x}.
\label{41}
\eeq
In the case of two co-planar detectors such an overlap is vanishing, quite independently of the relative distance and arm orientation. The overlap is vanishing also for detectors lying on two parallel planes, separated by an arbitrary distance. If, on the contrary, the arms of the interferometers are not lying on parallel planes, then the overlap of the common and differential modes is in general nonzero, even in the limit in which the spatial separation of the two central stations tends to zero, $\Da \vec x \ra 0$. 

For a simple illustration of this result we may consider a limiting configuration in which both interferometers are centered at the origin of the chosen frame. The arms of the first interferometer, 
$\hat u_1$ and $\hat v_1$, are aligned along the $x_1$ and $x_2$ axes, respectively, while the arms of the second interferometer, 
$\hat u_2$ and $\hat v_2$, are aligned along  $x_2$ and $x_3$, respectively. The two arms $\hat v_1$ and $\hat u_2$, in particular, are coaligned and coincident (see Fig. 3). The computation of the mixed, geodesic overlap function (\ref{41}), for such a configuration, leads to the result
\beq
 \ga_{+-}^{\rm geo}(p)= {\pi \over N} \left(p\over E\right)^2 \left( 2 - {7\over 15} {p^2\over E^2}\right).
\label{42}
\eeq
This overlap goes to zero as $(p/E)^2$ for $p \ra 0$: the corresponding signal induced by non-relativistic radiation is thus suppressed with respect to the relativistic radiation with $p=E$, but the suppression is lower than in case of two differential modes. 

\begin{figure}[t]
\begin{center}
\includegraphics[width=55mm]{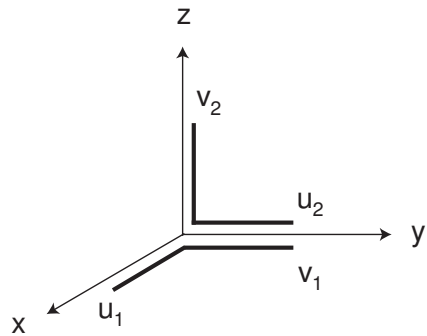}
  \end{center}
\caption{\sl Exampe of geometric arrangement with non-zero overlap of the common and differential mode.} 
\label{f3}
\end{figure}

Similar results can be obtained for the mixed (i.e., common-differential) overlap relative to a non-geodesic scalar, background, 
$ \ga_{+-}^{\rm ng}$. The result is nonzero only if the two interferometers are lying on non-parallel planes, like in Fig. 3.  In the non-geodesic case we recover however the usual, strong suppression for non-relativistic radiation, 
$ \ga_{+-}^{\rm ng} \sim (p/E)^4$ for $p \ra 0$. 

It should be stressed that the above results refer to the ideal situation in which the arms of the interferometers have exactly the same angular separation (in particular, they have been chosen to be orthogonal, according to Eq. (\ref{210})). In practice, however, we should also consider the case in which the angular separations of the arms are different. In that case, the mixed overlap may be nonzero even for co-planar detectors and for vanishing spatial separation,  
$\Da \vec x \ra 0$. 

\begin{figure}[t]
\begin{center}
\includegraphics[width=55mm]{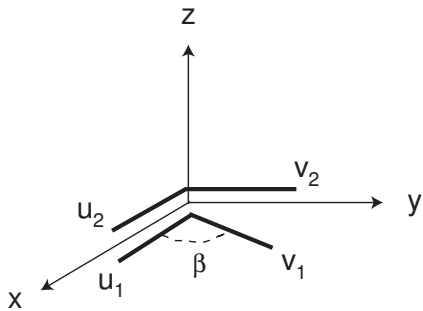}
  \end{center}
\caption{\sl Exampe of coincident interferometers with different angular separation of the two arms, and non-zero overlap of the common and differential mode.} 
\label{f4}
\end{figure}

In order to illustrate this point let us consider the (possubly realistic) experimental set up of Fig. 4, in which the arms of the first interferometer have angular separation $\b$ in the $(x_1,x_2)$ plane, while the arms of the second interferometer are orthogonal, in the same plane, with the arm $\hat u_2$ coincident and coaligned with $\hat u_1$:
\beq
\hat u_1 =\hat u_2=(1,0,0), ~~
\hat v_1=(\cos \b, \sin \b,0), ~~
\hat v_2=(0,1,0) 
\label{43}
\eeq
Such an experimental set-up could be simply realized, at least in principle, by adding a third, non-orthogonal arm to already existing interferometers. 

The response of the second, orthogonal interferometer is characterized by the scalar pattern functions already computed in in Eqs. (\ref{214}), (\ref{216}). For the second interferometer we obtain, from the definition (\ref{29}), a $\b$-dependent response,
\bea
F_-^{\rm geo}(\b)= &-& \left(p\over E\right)^2 \sin^2 \theta \Bigg[\cos 2\phi (\sin^2 \b) \nonumber \\
&& \qquad
- \sin 2 \phi(\sin \b \cos \b)\Bigg], \nonumber \\
F_+^{\rm geo}(\b)= 2 &-& \left(p\over E\right)^2 \sin^2 \theta \Bigg[1+\cos 2\phi (\cos^2 \b) \nonumber \\
&& \qquad
+ \sin 2 \phi(\sin \b \cos \b)\Bigg],
\label{44}
\eea
and similar expressions for the non-geodesic pattern functions. 

Let us now compute the mixed overlap $ \ga_{+-}^{\rm geo}$ at $d=0$. Such an overlap is identically zero, independently of $\b$, if we take the common mode of the orthogonal interferometer and the differential mode of the non-orthogonal one. We obtain a nonzero result in the opposite case, in which we overlap the common mode of the 
non-orthogonal interferometer with the differential mode of the orthogonal one:
\beq
 \ga_{+-}^{\rm geo}(p, \b)= {1\over N} 
 \int_{\Om_2} d^2 \n F_{1 +}^{\rm geo}(\b) 
F_{2 -}^{\rm geo}
={16 \pi \over 15 N} \left(p\over E\right)^4 \cos^2 \b. 
\label{45}
\eeq
Also in this, case, however, we recover the strong, non-relativistic suppression factor $(p/E)^4$. 

\section{Signal-to-noise ratio for non-relativistic backgrounds}
\label{V}

To discuss the sensitivity of the common mode to a scalar massive background we need to complete the computation of the 
signal-to-noise ratio ($SNR$), defined by \cite{1}:
\beq
SNR= \langle S \rangle /\Da S, ~~~~~~ (\Da S)^2= \langle S^2 \rangle
-\langle S \rangle^2
\label{51}
\eeq
(the brackets denote the optimized correlation over a time $T$, according to Eq. (\ref{31})). For uncorrelated noises, much larger than the physical strains, the variance $\Da S$ turns out to coincide with that obtained in the case of massless tensor radiation \cite{1}, 
\beq
(\Da S)^2= \langle S^2 \rangle
={T\over 4} \int_{-\infty}^{+\infty}d \nu~ P_1(|\nu|) P_2(|\nu|) |Q(\nu)|^2,
\label{52}
\eeq
where $P_i(|\nu|) $ are the so-called noise-power spectra of the two detectors, satisfying 
\beq
\langle n_i(t) n_i(t')\rangle= {1\over 2}
\int_{-\infty}^{+\infty}d \nu P_i(|\nu|)e^{-2\pi i \nu(t-t')}.
\label{53}
\eeq

Using the correlated signal (\ref{39}), and switching the momentum integration to the frequency domain $|\nu| \geq \bm$ through the standard relations $|\nu|= |E|= (p^2+\bm^2)^{1/2}$, $dp= (\nu/p) d\nu$, we can then express the SNR as 
\bea
&&
(SNR)^2={ \langle S \rangle^2 \over (\Da S)^2}= T\left(3 H_0^2\over 8 \pi^3\right)^2 {(Q,A)^2\over (Q,Q)}, 
\nonumber\\
&&
A={\ga_\pm (\xi)\Om (\xi)\over
|\nu|  \xi^2 P_1(|\nu|) P_2(|\nu|)} [\theta(\nu-\bm)+\theta(-\nu-\bm)],\nonumber\\
&&
\xi(\nu)= \sqrt{\nu^2-{\bm}^2}, 
\label{54}
\eea
where $\theta$ is the Heaviside step function, and the round brackets denote the following scalar product \cite{1}:
\beq
(A,B)= \int_{-\infty}^{+\infty}d \nu~A^\ast(\nu) B(\nu)  P_1(|\nu|) P_2(|\nu|).
\label{55}
\eeq
We have omitted, for simplicity, the indices ``ng" and ``X" because the above result generally applies both to the geodesic and non-geodesic response of the detector. The maximal value of the ratio (\ref{54}) obviously corresponds, like in the graviton case \cite{1}, to the choice of a filter function $Q$ proportional to $A$. The optimized value of $SNR$, written as an integral over the positive momentum domain, is finally given by (see also \cite{6})
\bea
&&
SNR= {3N\sqrt T H_0^2\over 8\pi^3} (A,A)^{1/2} =
\nonumber\\
&&
 ={3NH_0^2\over 8\pi^3}\Bigg[2\,T \,\int_{0}^{\infty}{dp\over  p^3\,(p^2+{\bm}^2)^{3/2}}
\nonumber\\
&& \qquad
\times
\frac{\ga_\pm^2(p)\,\Om^2(p)}
{P_1(\sqrt{p^2+{\bm}^2})\,P_2(\sqrt{p^2+{\bm}^2})}\Bigg]^{1/2}. 
\label{56}
\eea

It must be noticed, at this point, that the typical noise power spectra $P_i$ of present detectors are characterized by a minimum around a relatively narrow frequency band  $\Da \nu_0$, centered around $\nu_0$: outside that band the noises significantly increase,  and the ratio (\ref{56}) becomes negligible, $SNR \ra 0$. This implies that the detectors are only sensitive to spectral energy densities $\Om(p)$ falling within the resonant band $\Da \nu_0$. 

For a massive background, as already noticed in \cite{5}, there are in principle three possibilities. If $\bm \gg \Da\nu_0$ then the noises $P_i$ are always outside the sensitivity band, because $P_i(\nu)= P_i 
(\sqrt{p^2+{\bm}^2}) \gg P_i(\Da\nu_0)$ for all modes $p$, and there is no detectable signal.  If, on the contrary, $\bm \ll\Da\nu_0$, then the sensitivity band of the spectrum may possibly overlap with the relativistic branch of the spectrum (when $p\simeq \nu \sim \Da\nu_0$), but the non-relativistic branch $p<\bm$ always corresponds to a very high noise $P_i(\bm) \gg P_i(\Da\nu_0)$, and to a negligible signal. A resonant response to a non-relativistic spectrum can only be obtained in the third case, in which $\bm \sim \Da\nu_0$. In that case, indeed, the noises keep at their minimum for the whole non-relativistic branch of the spectrum,  because $P_i(\nu)=P_i(\sqrt{p^2+{\bm}^2})\simeq P_i(\bm)\sim P_i(\Da\nu_0)$ for all modes $0 \leq p \leq \bm$. It is just in that case that the signal due to the overlap of the common modes may show an interesting enhancement. 

In order to illustrate this possibility we may consider a non-relativistic spectral distribution with a peak and sharp cut-off at $p=\bm$. For instance \cite{6}
\beq
\Om(p)=\Om_0\left(p\over m\right)^\da, ~~~~~ p \leq \bm, 
~~~~~ \da >0,
\label{57}
\eeq
with $\Om_0=$ const. The spectral index $\da$ is positive to avoid the infrared divergence of the background energy density, $\int d \ln p~ \Om(p) < \infty$ (growing spectra are typical of relic backgrounds produced in the context of superstring cosmology and, in particular, pre-big bang inflation \cite{3,13}). The $SNR$ integral (\ref{56}) is always ultraviolet convergent because of the physical cut-off associated to the maximal energy mode amplified by inflation, and typical of any relic cosmic background. In the massless case the $SNR$ integral is also infrared convergent, because the limit $p=\nu \ra 0$ necessarily leads outside the resonant regime $\nu \sim \Da\nu_0$, and the noises blows up to infinity. In the massive case, on the contrary, $P_i(\nu) \ra P_i(\bm)=$ const when $ p \ra 0$, so that the infrared behavior of the $SNR$ integral is controlled by $\ga(p)$ and $\Om(p)$. For the given spectrum (\ref{57}), in particular, we are lead to the following integral 
\beq
I_\pm= \int_0^{\bm} dp~ p^{2\da-3} \ga_\pm^2(p).
\label{58}
\eeq

If we consider the overlap of two differential modes then $\ga_-(p)\sim (p/E)^4$ as $p \ra 0$ \cite{6}, and the above integral is always convergent for any growing spectrum. The same is true for the overlap of the non-geodesic response of two common modes, $\ga_+^{\rm ng}(p)$ (see Eq. (\ref{311})). For the geodesic response of the common mode, however, we have $\ga_+^{\rm geo}(p) \ra$ const for $p \ra 0$; as a consequence, the integral becomes
\beq
I_+^{\rm geo}\sim \int_0^{\bm} d p ~p^{2\da-3} \sim \left[p^{2(\da
-1)}\right]_0^{\bm}, 
\label{59}
\eeq
and is always divergent for flat enough spectra, $\da <1$, if extended down to $p=0$ (a similar effect is also found for the overlap of the monopole modes of two spherical, resonant mass detectors \cite{12}). Like in the case of spherical detectors such a divergence is unphysical, however, being removed by the fact that the physical observation time $T$ is not infinite, and is then associated to a {\em minimum} (non-zero) resoluble frequency interval, $\Da \nu = \Da E \gaq T^{-1}$. In the non-relativistic limit in which $\nu = E \simeq \bm + p^2/2 \bm$ this uncertainty condition defines a minimum momentum scale  \cite{12}
\beq
p\gaq p_{\rm min}= \left(2 \bm /T\right)^{1/2},
\label{510}
\eeq
to be used as the effective lower bound in the $SNR$ integration (modes with $p<p_{\rm min}$ cannot be resolved, and are to be included into the constant background over which scalar fluctuations are propagating, without contributing to the signal \cite{12}). 

For the geodesic response of two common modes, and for flat enough spectra, for which the momentum integral is dominated by the lower bound $p= p_{\rm min}(T)$, the $SNR$ thus acquires an anomalous dependence on the observation time $T$. Using the zero momentum limit $\ga_+^{\rm geo}(0)= 16\pi/N$ of Eq. (\ref{311}), using the spectrum (\ref{57}), and assuming that the two spectral noises $P_1, P_2$ are nearly equal to the same constant $P(\bm)$ for $p$ ranging from $0$ to $\bm$, we can obtain  the estimate (modulo unimportant numerical factors)
\bea 
(SNR)_+^{\rm geo} &\sim& \left(H_0\over \bm\right)^2
{\Om_0\over \bm ~P(\bm)}(\bm~ T)^{1-\da/2}, ~~ \da <1,
\nonumber \\
&\sim &\left(H_0\over \bm\right)^2
{\Om_0\over \bm ~P(\bm)}(\bm~ T)^{1/2}, ~~~ \da >1.
\label{511}
\eea
For $\da=1$ there is only a logarithmic correction to the usual time-dependence $T^{1/2}$. 

The result obtained for $\da>1$ is also valid for the non-geodesic response of two common modes, and for the geodesic and non-geodesic response of two differential modes \cite{6,11}, for any $\da>0$. We can see, therefore, that the signal associated to the geodetic response of the common mode grows faster with $T$ than in all other cases, provided the stochastic background is characterized by a sufficiently flat spectral distribution of its energy density. A typical observation time $10^7$ sec, for instance, leads to a relative enhancement  of $10^{7(1-\da)/2}$, for all spectra with $\da<1$.

\section{Conclusion}
\label{VI}

In this paper we have computed the angular pattern functions for the response of the common mode of  interferometric antennas to a stochastic background of massive scalar waves. We have taken into account both the direct (non-geodesic) interaction of the detector with the gradients of the scalar field background, and the indirect (geodesic) interaction with the scalar part of the metric fluctuations. The obtained results have been used to compute the overlap functions between the common modes, and also between the common and differential modes, of two independent detectors. 

We have found that the signal-to-noise ratio induced by a stochastic  background of non-relativistic particles is in general suppressed by the factor $(p/E)^4$, unless we consider the correlation of two common modes, geodesically coupled to the background. In that case the overlap function approaches a constant in the non-relativistic regime, and if the scalar spectrum is flat enough (so that the correlation integral  becomes dominated by its lower limit) then the signal-to-noise ratio grows faster with the correlation time $T$ than in all other cases. By extending the observation time it becomes possible, in principle, to enhance the signal with respect to the corresponding one  produced by the differential mode, possibly compensating the higher level of noise expected to affect the experimental analyses of the common mode data.

It is a pleasure to thank Carlo Ungarelli for a careful reading of the manuscript, and for many helpful suggestions. We are also grateful to Leonardo Angelini for his precious help in the preparation of  the angular plots displayed in this paper.

\end{document}